\documentclass[conference,10pt]{IEEEtran}
\IEEEoverridecommandlockouts
\usepackage{amsmath,amssymb,amsfonts}
\usepackage{algorithmic}
\usepackage{graphicx}
\usepackage{textcomp}
\usepackage{multirow}
\newcommand{\ie}{{\em i.e.,}\xspace}

\newcommand{\BfPara}[1]{\vspace{1mm}{\noindent\bf#1.}\xspace}
\usepackage{tikz}
\usepackage{booktabs}
\usepackage{balance}

\newcommand{\citet}[1]{\citeauthor{#1} \shortcite{#1}}

\usepackage{tikz}
\usepackage{amsmath}

\usepackage{hyperref}

\newcommand{\etal}{{\em et al.}}

\usepackage{comment}
\usepackage{blindtext}
\usepackage{booktabs} 
\usepackage{amsfonts}
\usepackage{amsmath} 
\usepackage{booktabs}
\usepackage{multirow}
\usepackage{graphics}
\usepackage{subfigure}
\usepackage{xcolor}
\usepackage{enumitem}
\usepackage{xspace}
\usepackage{makecell}
\newcommand{\vs}[1]{{\vspace{#1mm}}}

\hypersetup{
	colorlinks,
	urlcolor=blue,
	linkcolor=red,
	citecolor=green,
	bookmarksnumbered
}

\usepackage{multirow}
\def\BibTeX{{\rm B\kern-.05em{\sc i\kern-.025em b}\kern-.08em
    T\kern-.1667em\lower.7ex\hbox{E}\kern-.125emX}}
\begin{document}

\title{Hiding in Plain Sight: A Measurement and Analysis of Kids' Exposure to Malicious URLs on YouTube}

\author{\IEEEauthorblockN{Sultan Alshamrani} 
\IEEEauthorblockA{{\sl Department of Computer Science}\\
{\sl University of Central Florida}\\
\href{mailto:salshamrani@knights.ucf.edu}{salshamrani@knights.ucf.edu}}
\and 
\IEEEauthorblockN{Ahmed Abusnaina} 
\IEEEauthorblockA{{\sl Department of Computer Science}\\
{\sl University of Central Florida}\\
\href{mailto:ahmed.abusnaina@knights.ucf.edu}{ahmed.abusnaina@knights.ucf.edu}}
\and 
\IEEEauthorblockN{David Mohaisen} 
\IEEEauthorblockA{{\sl Department of Computer Science}\\
{\sl University of Central Florida}\\
\href{mailto:mohaisen@ucf.edu}{mohaisen@ucf.edu}}
}

\maketitle

\begin{abstract}
The Internet has become an essential part of children's and adolescents' daily life. Social media platforms are used as educational and entertainment resources on daily bases by young users, leading enormous efforts to ensure their safety when interacting with various social media platforms. In this paper, we investigate the exposure of those users to inappropriate and malicious content in comments posted on YouTube videos targeting this demographic. We collected a large-scale dataset of approximately four million records, and studied the presence of malicious and inappropriate URLs embedded in the comments posted on these videos. 
Our results show a worrisome number of malicious and inappropriate URLs embedded in comments available for children and young users. In particular, we observe an alarming number of inappropriate and malicious URLs, with a high chance of kids exposure, since the average number of views on videos containing such URLs is 48 million. When using such platforms, children are not only exposed to the material available in the platform, but also to the content of the URLs embedded within the comments. This highlights the importance of monitoring the URLs provided within the comments, limiting the children's exposure to inappropriate content.
\end{abstract}


\section{Introduction}
The influence of social media on the intellectual and emotional well-being of children and adolescents has been the sole focus of many studies recently, with social media being a central daily activity in their lives \cite{GasserCML12}.
Among different platforms, YouTube is the most popular video-sharing platform and is commonly used by children as an alternative to traditional TV, and as a source of entertainment and educational materials alike. 
A recent study by Smith~\etal~\cite{PewInternet2} reported that 81\% of U.S. parents allow their children to use YouTube for entertainment as an activity. Another study~\cite{familyzone} shows that children under the age of eight spent 65\% of their time on the Internet using YouTube.
Therefore, researchers have spent enormous efforts understanding the age-appropriate experience of children and adolescents when using YouTube, and have shown that inappropriate contents---such as contents with sexual hints, abusive language, graphic nudity, child abuse, horror sounds, and scary scenes---are common, with promoters for such contents targeting this demographic \cite{kaushalSBKP16,TahirASAZW19}.

To ensure the safety of young (children and adolescents) users, it is important to study their exposure to inappropriate material presented on YouTube including visual, audio, and written content. Even when watching videos from trusted family-friendly channels, the written contents, e.g., user comments, might contain inappropriate content that could influence their offline behavior.
Ensuring the safety of comments posted on kids' videos is not only limited to the toxic language being used in those comments~\cite{AlshamraniAAM20}, it also should include the detection of inappropriate and malicious contents~\cite{MohaisenAM15}. Malicious URLs embedded within the comments are one of the main threats studied in the Internet security community~\cite{SpauldingPKM18,ThomasM14,ManadhataYRH14,Mohaisen15,KosbaMWTK14,WestM14,ChoiAAWCNM19}, and are often used as a stepping stone to launching more advanced attacks, such as phishing, malware injection, and drive-by-download attacks~\cite{SaadAM19}. Attackers create a malicious web page that can be used to perform different modes of attack then send the malicious URLs to other people, or post them on public forums in the hope that others will click on them to initiate the attack procedure~\cite{ZhangM0LX19}. Once victims click on the malicious URL, for example, they will be taken to that malicious web page without notice, especially if such users are kids and adolescents, without the appropriate level of security awareness~\cite{EsheteVWZ13}.

Our study explores measuring the exposure of children and adolescents to age-inappropriate and malicious external URLs in YouTube comments posted on videos of the top-200 children's shows~\cite{Ranker}. This task is challenging for several reasons. First, studying comments on children's videos requires manually collecting channels and shows targeting this demographic, knowing YouTube categories are not established by age-group but rather by the topic they convey. Second, assigning age groups to the collected videos can be daunting in measuring exposure by separate groups.
To address these challenges, we build a large collection of YouTube comments on children-oriented videos for the top 200 shows categorized by different age groups~\cite{CommonSenseMedia}. 
We then extract the URLs within the collected comments, and analyze them, uncovering a large number of age-inappropriate and malicious URLs embedded in the comments posted on children YouTube videos, particularly, videos with a very high average number of views.

\begin{figure*}[t]
\centering
\begin{minipage}[t]{0.32\textwidth}
\includegraphics[width=0.99\textwidth]{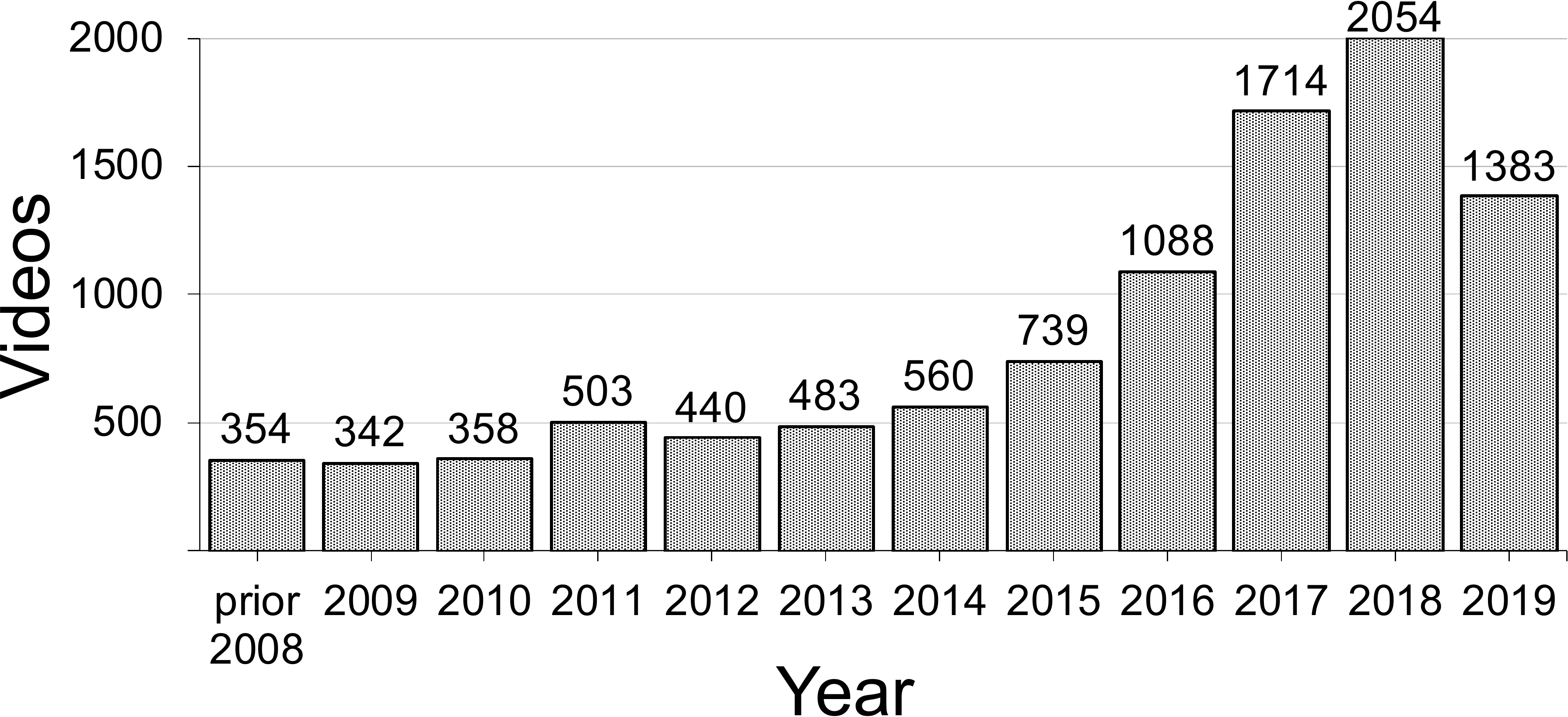}
\caption{The publish date distribution of the collected YouTube kids' videos.}
\label{fig:videos_years}
\end{minipage}
\hfill
\begin{minipage}[t]{0.32\textwidth}
\includegraphics[width=0.99\textwidth]{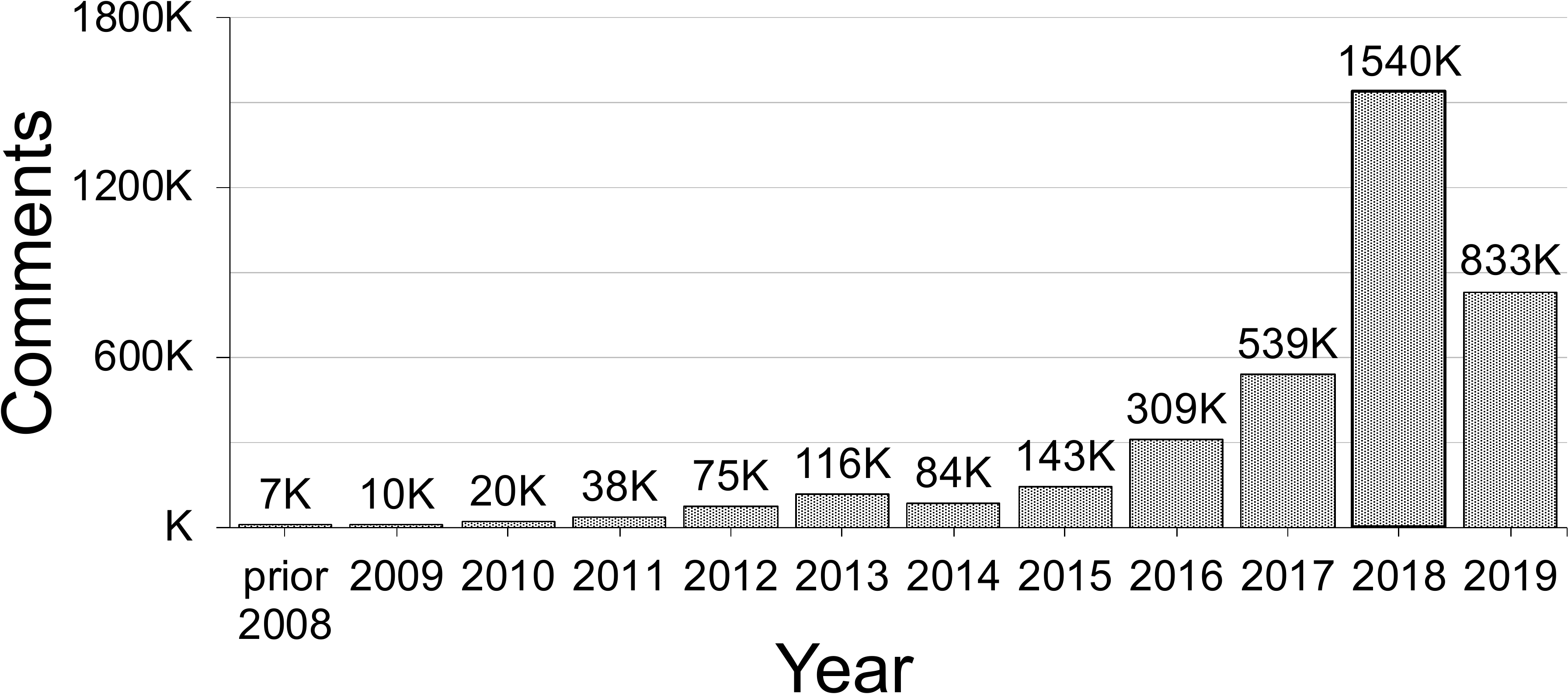}
\caption{The distribution of YouTube kids' videos comments over past years.}
\label{fig:comments_years}
\end{minipage}
\hfill
\begin{minipage}[t]{0.32\textwidth}
\includegraphics[width=0.99\textwidth]{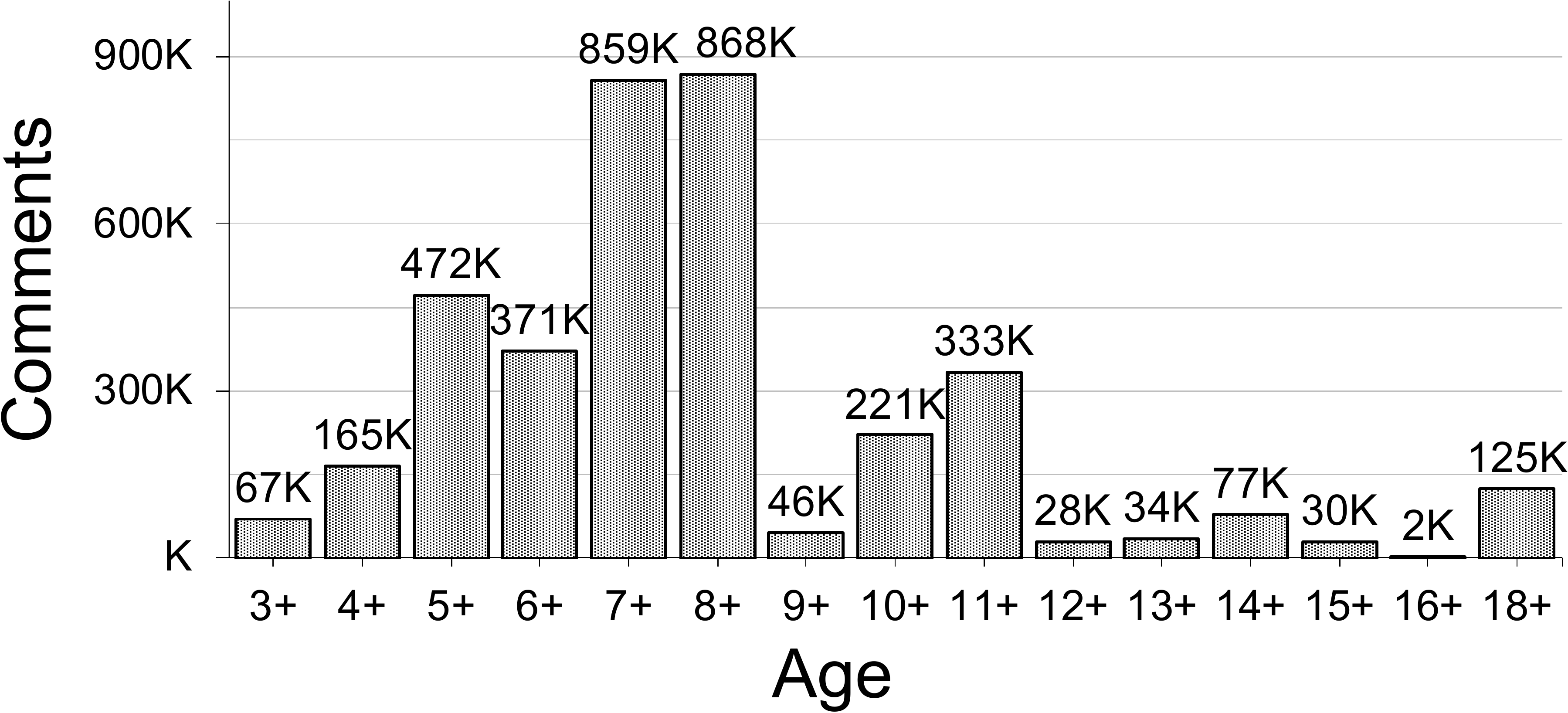}
\caption{The distribution of YouTube kids' comments over different ages.}
\label{fig:age_comments}
\end{minipage}
\end{figure*}

\BfPara{Contributions}
This work contributes to the space of measuring the exposure of children and adolescents to inappropriate and malicious URLs present in the kids' YouTube videos comments. We summarize our contribution as follows: 
\begin{itemize}\itemsep=1.5mm
    \item We collected a large-scale dataset of comments on children's YouTube videos from the top-200 ranked children shows. The list of shows, retrieved search results, categorization of shows by age group, and other aspects related to the data collection process are manually vetted. 
    \item We conduct an in-depth analysis of kids' exposure to URLs embedded in the comments as well as their interactions with comments and videos having the URLs. Among the collected dataset, we extracted 8,677 URLs. Further studying the URLs associated topics and audience interaction with inappropriate websites, such as illegal content and adult websites.
    \item We report on a sizable number of URLs appearing in the comments on the analyzed videos that are marked as malicious by VirusTotal. We explored the user interaction with such comments, and their popularity, uncovering the risks associated with the exposure to such content. 
\end{itemize}

\section{Related Work}
The increasing awareness of the effect of social media on children brought attention to how appropriate the provided contents are for children.  
Several studies explored the effect of social media on children, including O'Keeffe~\etal~\cite{OGCK2011}, in which, authors encouraged parents to understand and be aware of offline and online behaviors of their children such as cyber-bullying, privacy issues, sexting, and Internet addiction.

Studying the appropriateness of contents being presented to children on YouTube was first raised, to the best of our knowledge, by Kaushal~\etal~\cite{kaushalSBKP16}, who studied kids-unsafe contents and promoters. The authors provided a framework for detecting unsafe content using measures calculated on the video, user, and comment levels with an accuracy of 85.7\%.  
More recently, Tahir~\etal~\cite{TahirASAZW19} shows that even children-focused apps such as {\em YouTube Kids}, which are considered kids-safe platforms, are prone to compromise with inappropriate videos. 

Studying comments and user feedback, one of the earliest studies by Bermingham~\etal~\cite{BerminghamCMOS09} provided sentiment analysis of topics potentially serving a radicalizing agenda using a dataset of YouTube channels profiles and user comments.
Ezpeleta~\etal~\cite{EzpeletaIGMZ18} used mood analysis to improve spam filtering accuracy for comments. Comments on YouTube have been studied by Cunha~\etal~\cite{CunhaCP19} to analyze users' opinions on several aspects such as the quality of the video, YouTuber presence, and videos' contents. Poche~\etal~\cite{PocheJWSVM17} showed that studying users' comments on YouTube coding tutorials can help the channel owner providing better content, which enables achieving higher popularity as Figueiredo~\etal~\cite{FigueiredoABG14} showed. In particular, the quality and users’ perception of content facilitates popularity on YouTube. 

In this work, we focus on studying the URLs present in the kid's YouTube videos comments. While previous studies focused on analyzing the video, caption, and comments, it is of high importance to understand how appropriate and safe the URLs are, as children may access their content intentionally or accidentally. In particular, we focus on URLs that contain inappropriate topics for children, such as pornography, politics, weapons, and adult contents, and malicious URLs that are used to harm the users and their devices.

\begin{figure*}[t]
\centering
\begin{minipage}[t]{0.32\textwidth}
\includegraphics[width=0.99\textwidth]{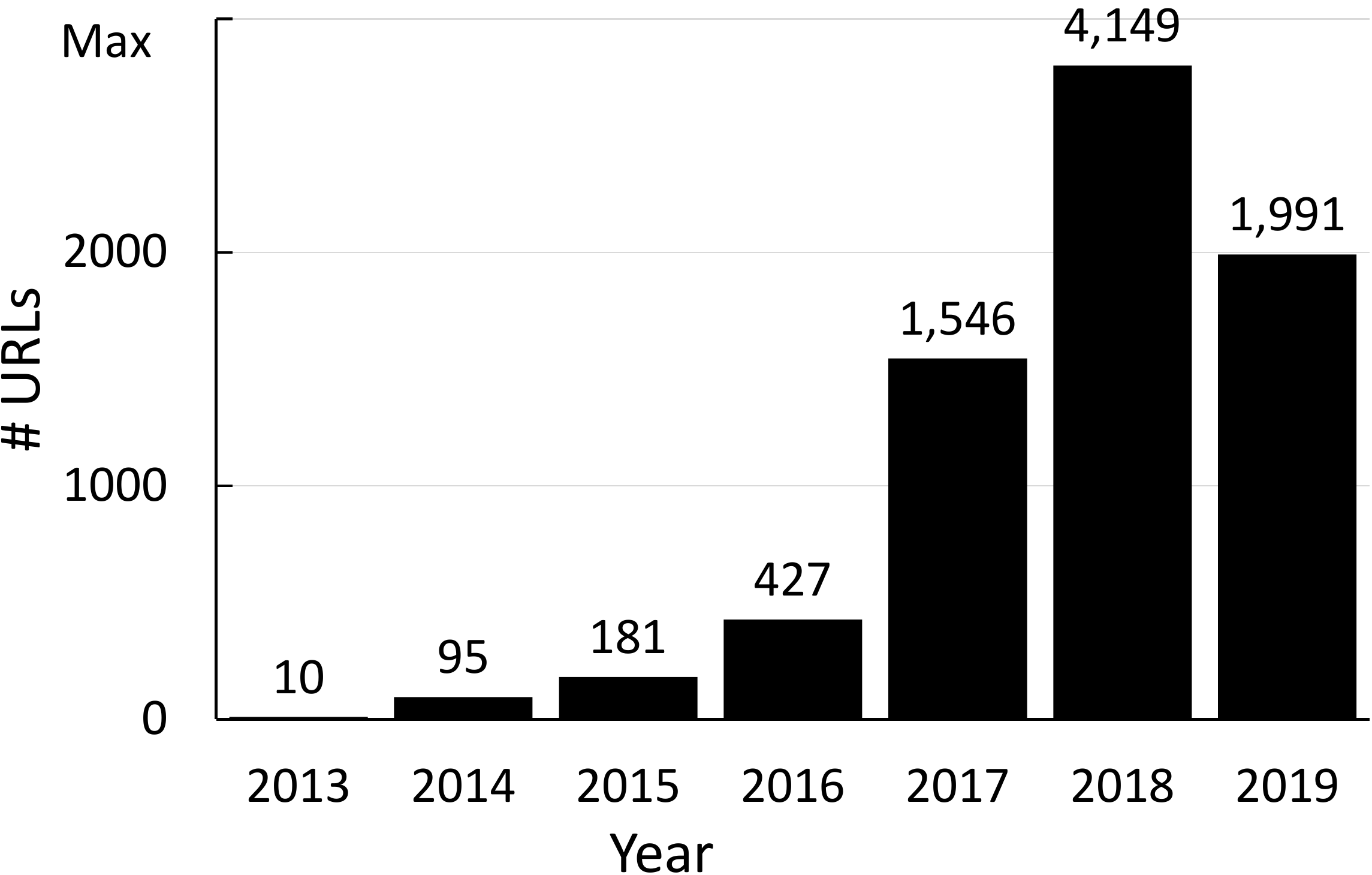}
\caption{The distribution of the collected URLs over the years.}
\label{fig:cate_years}
\end{minipage}
\hfill
\begin{minipage}[t]{0.32\textwidth}
\includegraphics[width=0.99\textwidth]{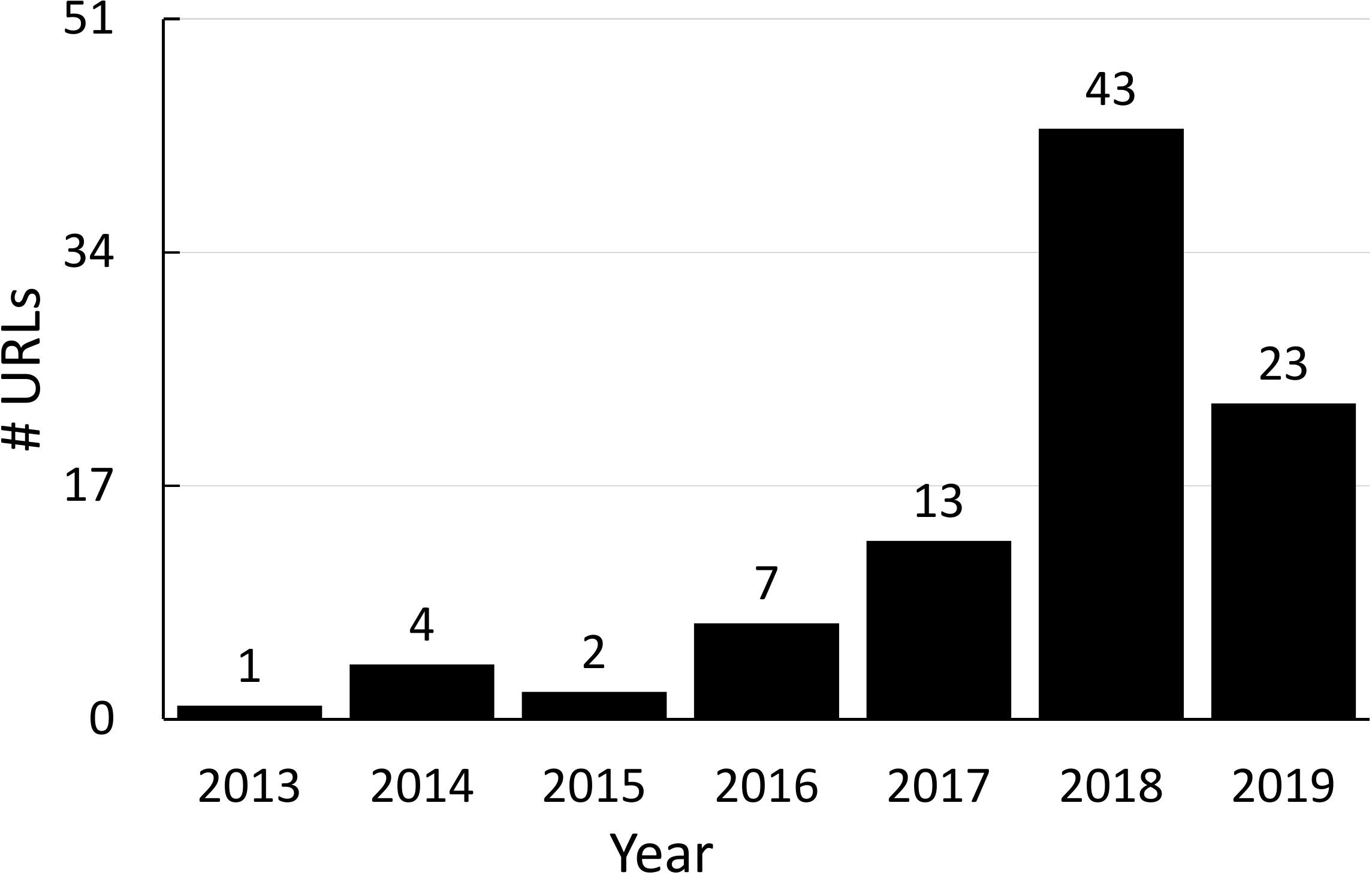}
\caption{The distribution of the inappropriate URLs over the years.}
\label{fig:8cate_year}
\end{minipage}
\hfill
\begin{minipage}[t]{0.32\textwidth}
\includegraphics[width=0.99\textwidth]{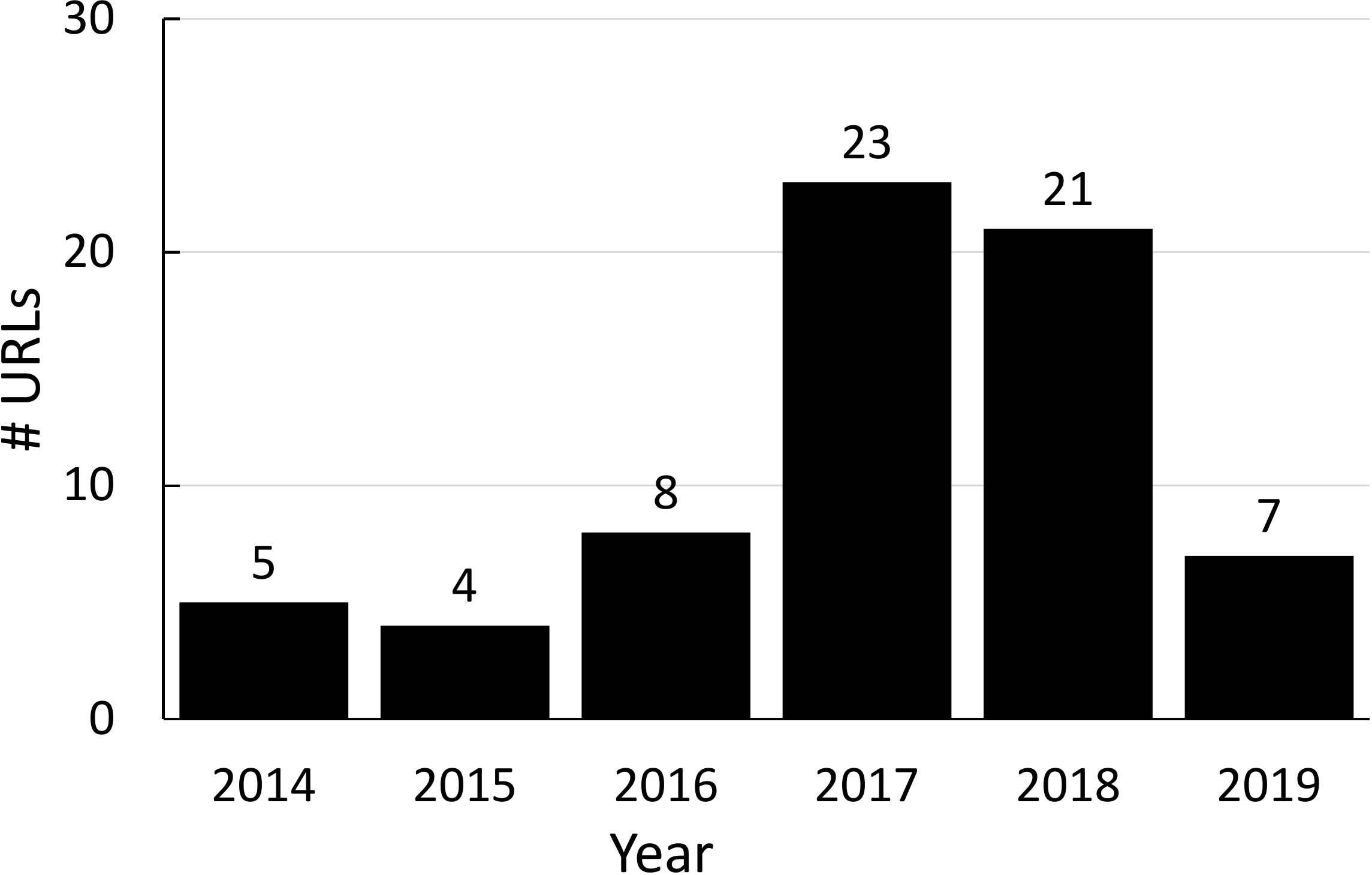}
\caption{The distribution of the Malicious URLs over the years.}
\label{fig:Malicious_urls_year}
\end{minipage}
\end{figure*}

\begin{figure}[ht]
\centering
\includegraphics[width=0.40\textwidth]{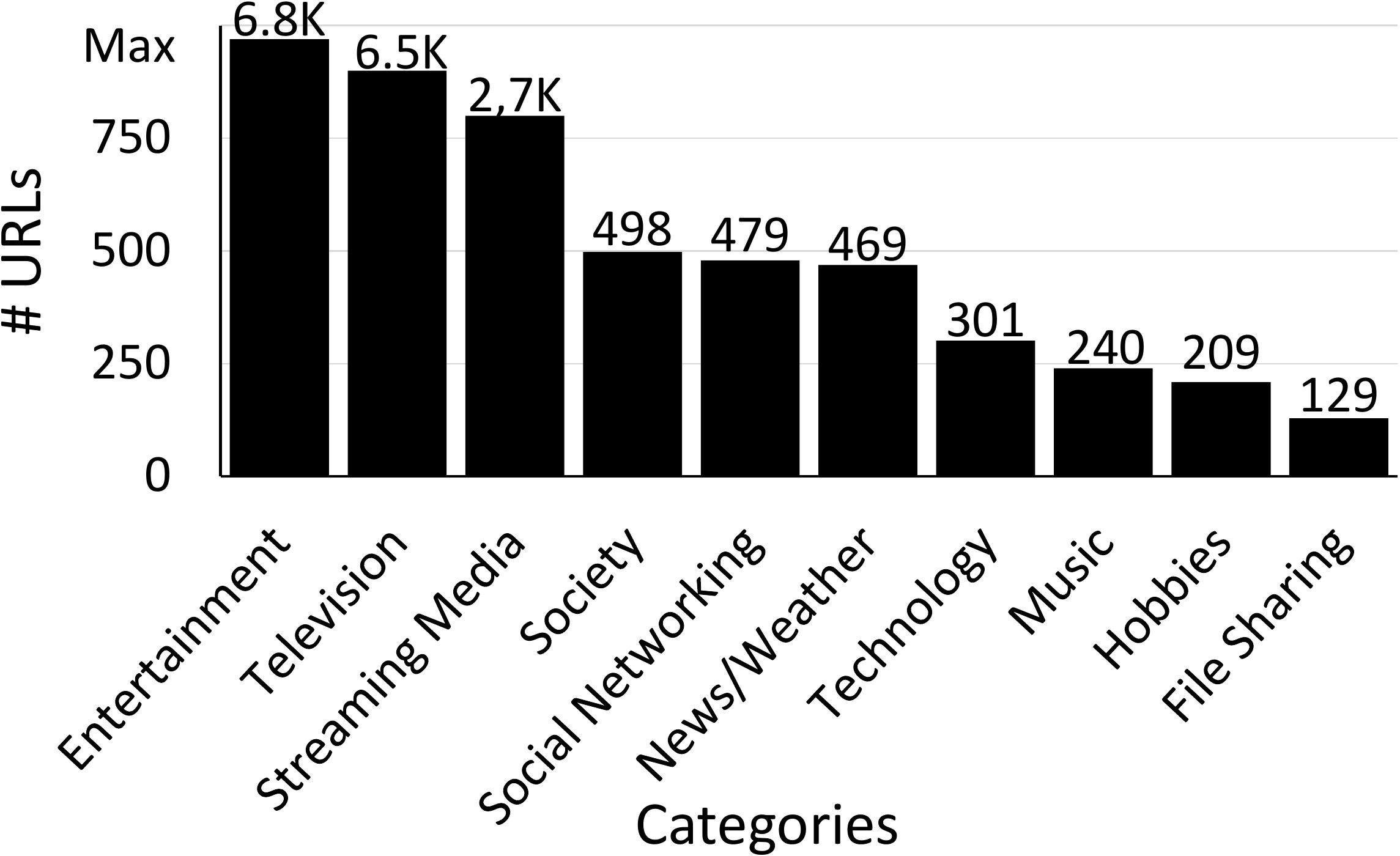}
\caption{The top 10 IAB Categories associated with the collected URLs.}
\label{fig:top_10cate}
\end{figure}

\section{Research Questions, Datasets, and Preliminary Analysis}

\subsection{Research Questions}
The main goal of this study is to answer various questions around the safety of contents generated specifically for kids and adolescents published on YouTube. The study is framed around several research questions, as follows:
\begin{itemize}[leftmargin=3mm]
    \item {\bf RQ1:} Are children and adolescents on YouTube kids videos exposed to external resources and content? We address this question by collect a sizable dataset of 3.7 million comments posted on roughly 10,000 YouTube kids videos, extracting 8,677 associated URLs embedded within the comments.
    \item {\bf RQ2:} How appropriate the URLs embedded in comments on videos targeting children and adolescents? We address this question by extracting the URL domain category (topic), and studying the appropriateness of the URL topic for children and adolescents.
    \item {\bf RQ3:} Are comments on videos targeting children and adolescents associated with malicious URLs? We address this question by forwarding the URL to a malicious website detection service, uncovering 68 malicious URLs embedded within the dataset.
    \item {\bf RQ4:} What kind of malicious intent do those malicious URLs serve? We address this question by associating the malicious URLs with three malicious attributes, including ``malicious'', ``malware'', and ``phishing'' websites.
    \item {\bf RQ5:} What are the likelihood of URLs posted on videos targeting this audience to engage with the audience? Address this question directly would involve information that is not explicitly published on YouTube. To address this question, however, we use contents from our data collection and inferences that can support this exploration. We defined two metrics to estimate the interaction with the URL by the audience, including video popularity, represented, and URL's comment popularity.
\end{itemize}

\subsection{Data Collection and Measurements}
The data used in this study include 3.7 million YouTube comments posted on roughly 10,000 children's videos, distributed over the period from January, 2005 until March, 2019.

\BfPara{Children's Shows}
We collected comments on videos of the top-200 children's shows based on Ranker~\cite{Ranker}, a website that relies on millions of users to rank a variety of media contents such as shows and films. 
The list of shows was originally made by Ranker TV, and has more than 1.2M votes on it from 32,500 registered users, and has been viewed by more than 400,000 viewers. The list has 380 kids' shows in which we have selected the top 200 shows. We then extended our list of children's shows from a Wikipedia list of cartoon shows.

\BfPara{Collection Approach} 
Using YouTube APIs, we extracted the top-50 videos of the search results on every show on our list. The search API allows us to retrieve video IDs, which we used to obtain statistics about each video such as the number of views, likes, dislikes, etc.
Then, we used YouTube Comments API to collect all comments from the videos. 
In total, we collected more than 3.7 million comments from 10,000 videos. 

\BfPara{Age-Appropriateness of Children's Shows} 
We defined age appropriateness as the adequate age group to be the subject of the show. 
Defining the age appropriateness for children's shows is challenging, since most shows do not specify the target age group. 
Therefore, we used  {\em Common Sense Media}~\cite{CommonSenseMedia}, a non-profit organization that provides education and advocacy to families on providing safe media for children, as the main source for defining the age group of the targeted children's shows.  
Using {\em Common Sense Media}, we retrieve the appropriate age group for most kids' shows on our list. 

A few shows did not appear in {\em Common Sense Media}, and for those we turned to IMDB\cite{imdb} to obtain the age group.  
Moreover, some shows have different versions, for different age groups, and for those we assigned the age group based on the most prevalent version in the YouTube search. 
Some kids show are assigned an age group based on their respective categories, e.g., Loony Tunes (a well-known collection of cartoons for age 7+).    
We manual inspected the age appropriateness for the retrieved top-50 results on each show to define non-kids contents and assigned them to the 17+ age group, which is the highest age group in our dataset.

\begin{figure*}
\centering
\begin{minipage}[t]{0.49\textwidth}
\centering
\includegraphics[width=0.90\textwidth]{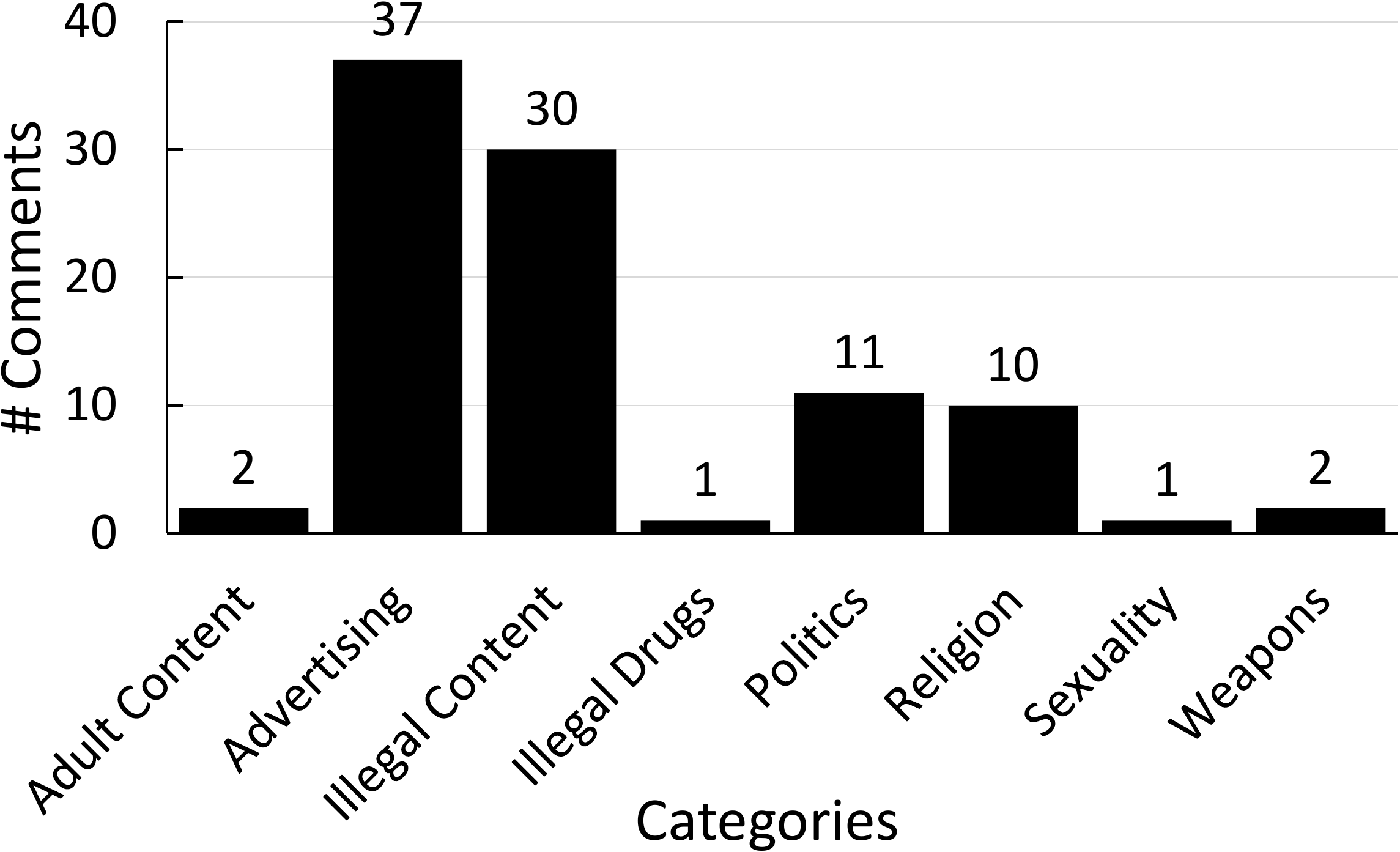}
\caption{The distribution of inappropriate URLs over different IAB Categories.}
\label{fig:urls_cate}
\end{minipage}
\hfill
\begin{minipage}[t]{0.49\textwidth}
\centering
\includegraphics[width=0.90\textwidth]{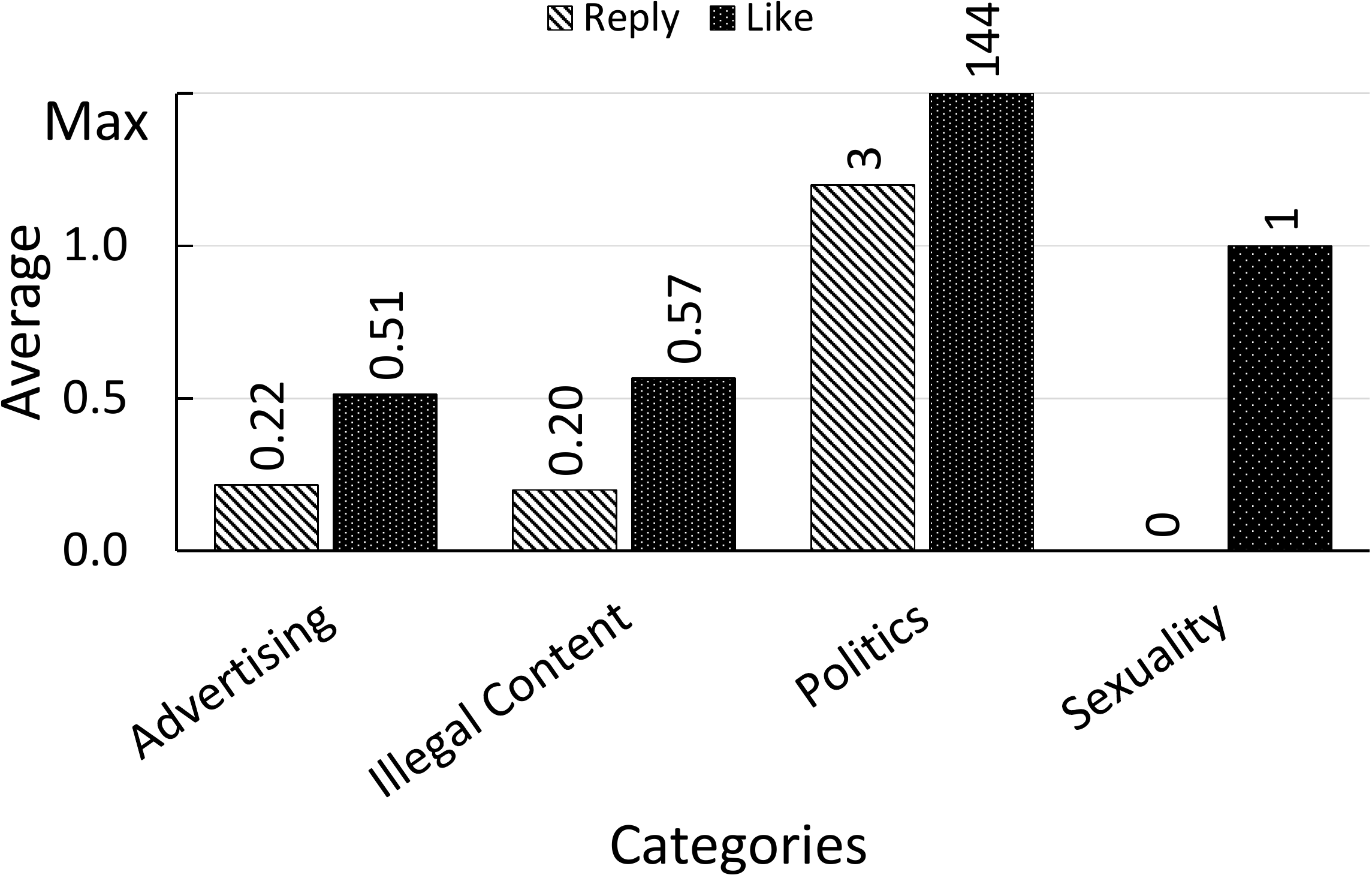}
\caption{Users' interactions with inappropriate URLs from different Categories.}
\label{fig:cate_interaction}
\end{minipage}
\end{figure*}

\BfPara{Data Statistics and Measurements}
The collected YouTube comments were posted by more than 2.5 million users on approximately 10,000 videos.   
These retrieved videos have an average viewers count of roughly 2.4 million views and an average comments count of 8,068 comments per video. 
Observing the publishing date of the videos in our collection, Figure~\ref{fig:videos_years} demonstrates the rapid increase in children's videos gained over the past few years. 
The figure shows an increase in popularity of five folds in ten years from 2008 (with 354 videos) to 2018 (with 2,054 videos).  
This rapid growth in popularity is observed through the first three months of 2019 with 1,383 videos included in our collection (by March 2019).

We note that the collection of YouTube videos is based on the relevance, and not the publishing date nor the view count. 
This is also the case when retrieving videos from the top-50 search result and querying the targeted shows.
We note that the search results do not always reflect popularity, although the top-ranked videos are often characterized by bursts of popularity \cite{FigueiredoBA11}.
Generally, a consistent trend is observed in the year-over-year increasing number of videos in our collection.

Similar patterns are observed with the number of comments from around 7,000 comments on videos prior to 2008 to more than 1.5 million comments on videos from 2018. 
This growth is steady through the first three months of 2019, per Figure~\ref{fig:comments_years}.
We also provide the distribution of comments across the age groups as shown in Figure~\ref{fig:age_comments} where most of the collected comments were posted on videos for kids between the age of 5 to 8 (a total of approximately 2.5 million comments). 

\subsection{URL Extraction}
The main focus of this study is to explore how appropriate are the contents of comments on YouTube kids videos. In particular, we focused on the URLs embedded in the comments, and investigated their potential risks on children. 
We used a regular expression to extract possible URLs within the comments. 
In the collected dataset, we extracted 8,677 URLs, associated with 1,628 videos.
Figure~\ref{fig:cate_years} shows the number of URLs extracted per year. Notice that there is an increasing trend of the number of URLs embedded in the comments, shedding the light on the importance of monitoring the content of the comment, particularly in children-oriented channels

\BfPara{URL Topic Categorization}
We extracted the topics associated with the embedded URLs to understand their effects on children's exposure to various contents.
In particular, we used Webshrinker~\cite{webshrinker}, a machine learning-powered domain data, and threat classifier, to obtain the Interactive Advertising Bureau (IAB) categorization of the domains of the URLs. 
To this end, we extracted 107 different categories associated with the URLs.
Figure~\ref{fig:top_10cate} shows the top ten categories associated with the URL. 
Note that a URL may be associated with one or more category, based on IAB categorization. 
Note also that entertainment, television, alongside with streaming media, were the most common categories within the URLs. 

\BfPara{Malicious URL Extraction}
In the context of videos targeting kids, the chance that the audience will blindly click on the URLs posted on videos is very high. 
Such a behavior may allow attackers to gain information or access to private resources on the victim's device.
To this end, we extracted all the URLs within the collected comments and checked whether the given URL is valid or not by accessing the website and checking the response of the HTML request. 
If the returned response status code is 200 (success), we then forward the URL to VirusTotal API~\cite{VirusTotal} to check whether it is benign or malicious as well as the URL's associated attributes. Those attributes include ``malicious'', ``malware'', and ``phishing''. Malicious websites contain exploits or other malicious artifacts, malware websites are used for malware distribution, and phishing websites are used for stealing users' credentials or private information.

\begin{figure*}
\centering
\begin{minipage}[t]{0.32\textwidth}
\includegraphics[width=0.99\textwidth]{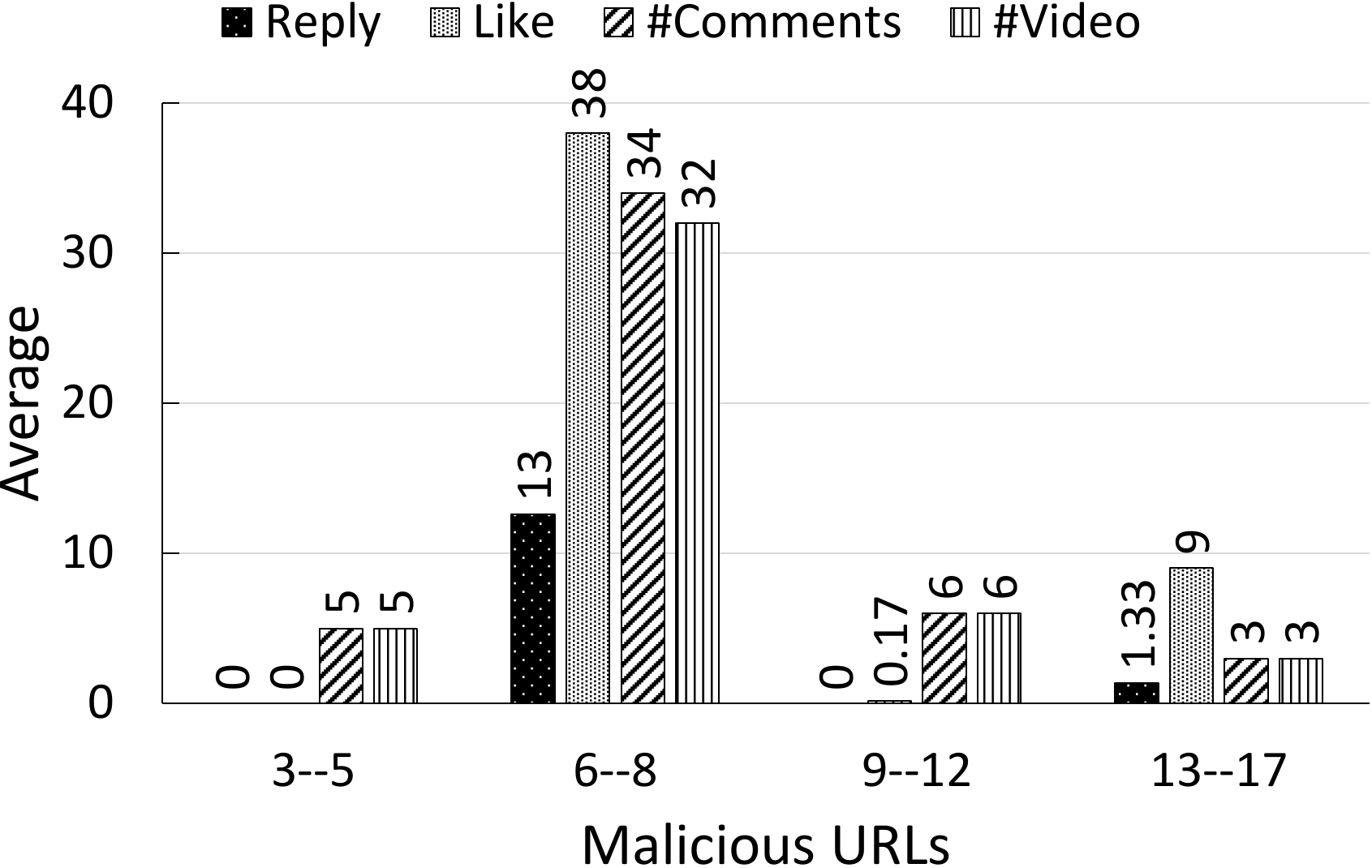}
\caption{Users' interactions with Malicious URLs for all age groups.}
\label{fig:Malicious urls_age}
\end{minipage}
\hfill
\begin{minipage}[t]{0.32\textwidth}
\includegraphics[width=0.99\textwidth]{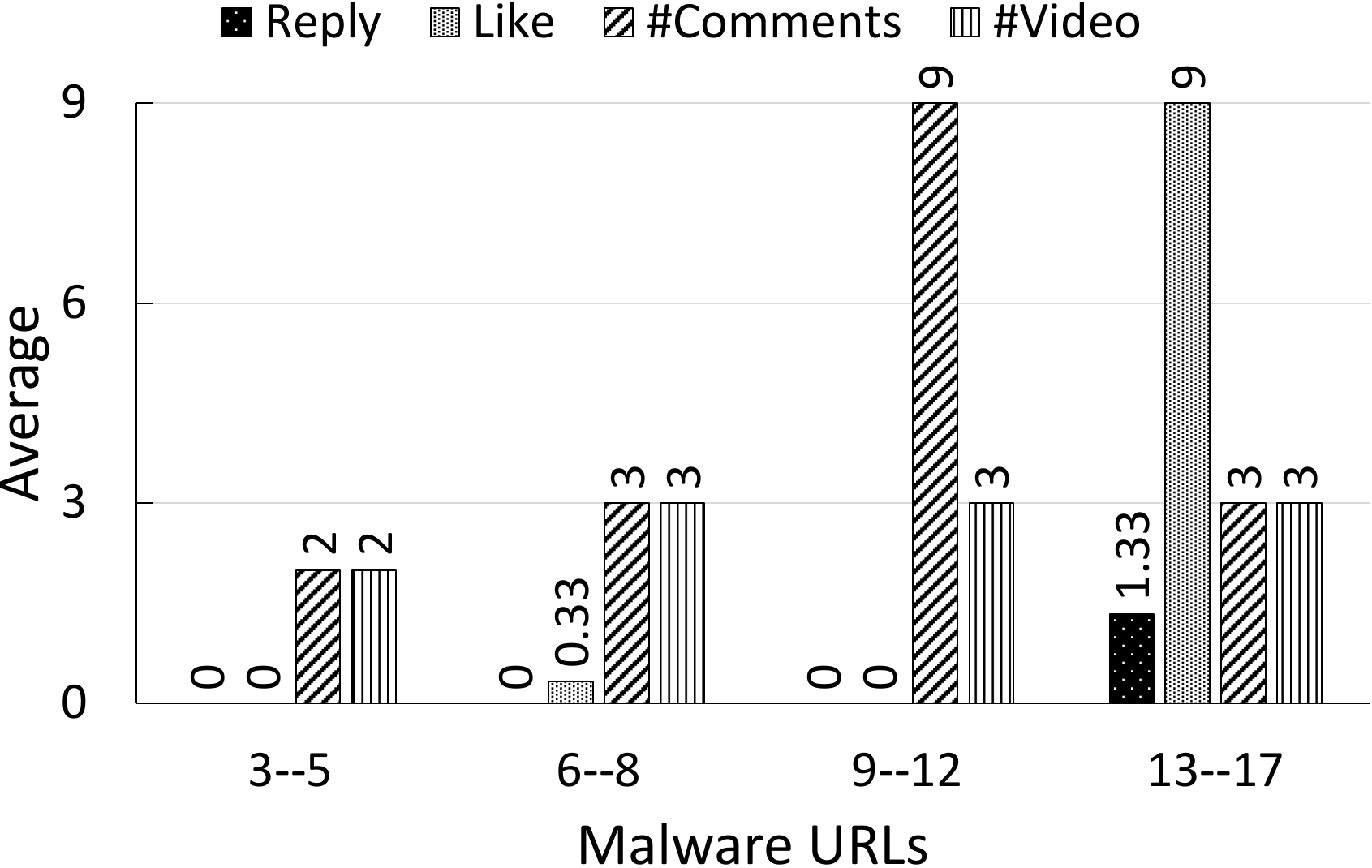}
\caption{Users' interactions with Malware URLs for all age groups.}
\label{fig:Malware urls_age}
\end{minipage}
\hfill
\begin{minipage}[t]{0.32\textwidth}
\includegraphics[width=0.99\textwidth]{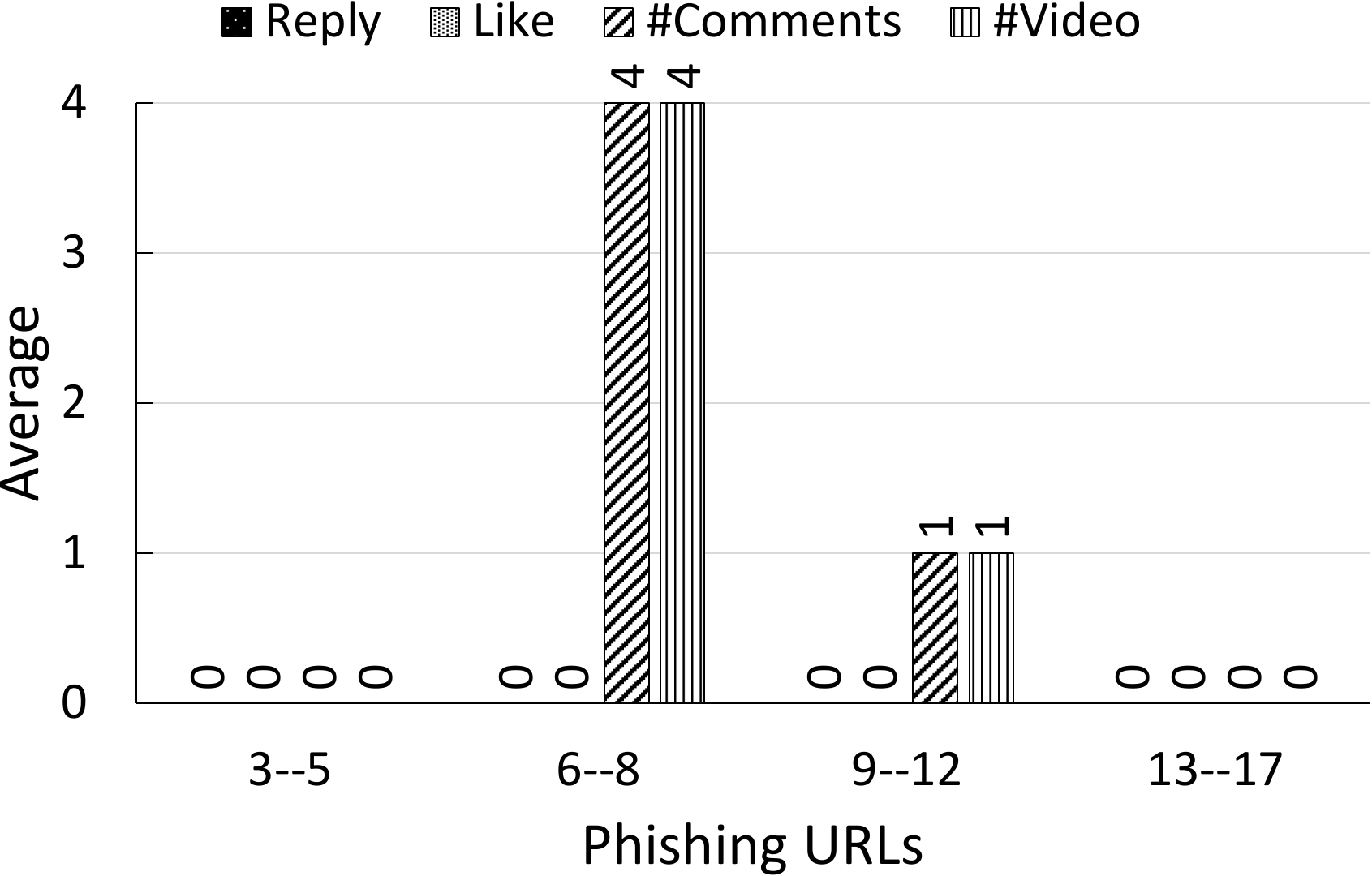}
\caption{Users' interactions with Phishing URLs for all age groups.}
\label{fig:Phishing urls_age}
\end{minipage}
\end{figure*}
\begin{table*}[]
\centering
\caption{Distribution of the detected malicious URLs over different age groups as well as the average number of viewers, likes, dislikes and replies for each age group. $^1$ Analysis including all videos and comments collected in the dataset.} \vs{3}
\label{tab:age_mal}
\scalebox{1.0}{
\begin{tabular}{l|c|c|c|c|c|c|c|c}
\Xhline{2\arrayrulewidth}
          & \multicolumn{5}{c|}{Videos with  Malicious URLs}                                              & \multicolumn{3}{c}{Comments with Malicious URLs}           \\
          \cline{2-9}
Age group & \#Videos & Avg\_comments & Avg\_viwers    & Avg\_likes  & Avg\_dislikes & \#Comments & Avg\_likes & Avg\_replies \\
\Xhline{2\arrayrulewidth}
3-5 &	7	& 11,218 &	232,743,621 & 	330,894	 &  125,944	&  7	& 0	& 0         \\
6-8 &	39	& 20,714 &	21,138,213 & 	211,227 &	14,122	& 41	&142 & 	10        \\
9--12 &10	& 24,395	& 42,455,109	& 206,904	& 18,647	&16	&0.06	&0         \\
13-17	& 4	&2,203	& 1,380,479&	14,563&	806&	4&	9&	1.33         \\
\Xhline{2\arrayrulewidth}
Total    &60	&18,958	&48,043,712	&211,297	&27,026	&68	&86	&6.44      \\
\Xhline{1\arrayrulewidth}
Overall$^1$      &9,996	&650	&2,106,472	&8,119	&1,230	&3,712,911	&9.87	&0.50      \\
\Xhline{2\arrayrulewidth}
\end{tabular}}

\end{table*}

\begin{table*}[]
\centering
\caption{Distribution of videos and comments containing different malicious URLs as well as the average number of viewers, likes, dislikes and replies for each malicious type. $^1$ Analysis including all videos and comments collected in the dataset.} \vs{3}
\label{tab:mal_type}
\scalebox{1.0}{
\begin{tabular}{l|c|c|c|c|c|c|c|c}
\Xhline{2\arrayrulewidth}
               & \multicolumn{5}{c|}{Videos with Malicious URLs}                                                      & \multicolumn{3}{c}{Comments with Malicious URLs}              \\
               \cline{2-9}
URLs Type & \#Videos & Avg\_comments & Avg\_viwers & Avg\_likes & Avg\_dislikes & \#Comments & Avg\_likes    & Avg\_replies \\
\Xhline{2\arrayrulewidth}
malicious site&	47&	10,017&	46,061,532&	136,621&	24,234&	49&	120&	8.94         \\
malware site&	8&	26,288&	51,075,237&	284,887&	28,923&	14&	0.07&	0         \\
phishing site&	5&	91,286&	61,825,765&	795,507&	50,234&	5&	0&	0         \\
\Xhline{2\arrayrulewidth}
Total	&60	&18,958	&48,043,712	&211,297	&27,026	&68	&86	&6.44 \\
\Xhline{1\arrayrulewidth}

Overall$^1$      &9,996	&650	&2,106,472	&8,119	&1,230	&3,712,911	&9.87	&0.50      \\
\Xhline{2\arrayrulewidth}
\end{tabular}}

\end{table*}

\section{URL Content Analysis}
This study investigates how appropriate the URLs embedded in the comments on YouTube kids' videos.
As previously mentioned, the audience may intentionally or accidentally access the content of the URLs, highlighting the importance of understanding the content and its effect on the children.
While it is not possible to know how many users accessed the URLs, we defined two metrics to estimate the prevalence and use of the URL by the audience, including 1) video popularity, represented by the number of views, likes, and comments on the video including the URL, and 2) comment popularity, defined as the likes and replies on the comment containing the URL. While the latter does not necessarily capture the context of the engagement, it captures the level of engagement as a magnitude, which is essential in capturing users' exposure.

\subsection{Kids Exposure to Inappropriate Topics}
Within the 107 IAB extracted topics, eight topics are highly inappropriate for children, including ``\textit{Adult Content}", ``\textit{Advertising}", ``\textit{Illegal Content}", ``\textit{Illegal Drugs}", ``\textit{Politics}", ``\textit{Religion}", ``\textit{Sexuality}", and ``\textit{Weapons}". Note that other topics may not be appropriate for children, however, we only considered the most obvious topics that are directly inappropriate for children to be exposed to. 

Figure~\ref{fig:8cate_year} shows the number of URLs associated with inappropriate topics. In total, 94 URLs were classified as inappropriate, with an increasing trend in such URLs over the years, with three folds increase between 2017 and 2018. This indicates the risk of children's exposure to worrisome content that is not appropriate for their age. Figure~\ref{fig:urls_cate} shows the distribution of the URLs among different inappropriate topics. While topics such as ``\textit{Advertising}" and ``\textit{Illegal Content}" are popular within the URLs, with 71.27\% of the total URLs associated with these two categories. 

In addition, it is important to understand the users' interaction with the URLs' comments. Figure~\ref{fig:cate_interaction} shows the number of likes and replies associated with four different inappropriate categories. Note that the other categories were excluded as the users did not interact with their comments. As shown, comments with political URLs have on average three replies, and 144 likes, which is abnormal given the video/channel targeted audience. In general, the inappropriate URLs within the YouTube kid's comments are on the rise, leading to a potential risk of kids' exposure to their content.

\subsection{Kids Exposure to Malicious URLs}
Measuring kids' exposure to malicious URLs by different age groups,
Figure~\ref{fig:Malicious_urls_year} shows over years number of malicious URLs embedded in the comments. Similar to the inappropriate topics, the number of comments with malicious URLs is increasing over the years. 
Note that our collected dataset only includes the first three months of 2019. 
Figure~\ref{fig:Malicious urls_age} highlights the interaction of each age group with malicious URLs, where kids from the age of 6 to 8  have the highest interaction with malicious URLs, represented as the average number of replies, likes, comments, and videos. 
Furthermore, we studied the kids' interaction with malware URLs, as shown in Figure~\ref{fig:Malware urls_age}. Here, the age groups 9-12 and 13-17 show the highest interaction with malware URLs, represented with the likes and comments on the mentioned comments. 
Similarly, Figure~\ref{fig:Phishing urls_age} shows kids' interaction with phishing URLs. 
Note that only two age groups (\ie 6-8 and 9-12) include phishing URLs, however, the users did not interact with their comments.   

In more detail, Table \ref{tab:age_mal} shows that videos with malicious URLs embedded in their comments have high users' interaction and engagement, which can be seen in the average number of views, comments, likes, and dislikes. There are a total of 41 videos with malicious URLs embedded in their comments, with an average of more than 46 million viewers. Based on that analysis, we can safely consider these videos as popular, which would attract more users. Moreover, the videos with malicious URLs targeting kids from the age of 3 to 5 have the highest average number of viewers, with more than 200 million views, followed by the age group 9-12, with around 42.4 million views.

Table \ref{tab:mal_type} lists the three types of malicious websites with the number of videos and comments containing each of the malicious URLs, as well as general statistics that show users' interactions with each of them. We can see that videos with malware sites URLs have an average number of viewers of more than 51 million views, which makes the possibility for a large number of people getting affected by malware much higher. The results also show that there are more than 61 million viewers of the videos with phishing URLs embedded in their comments, which is also an alarming finding in itself, since a higher number of viewers increases the likelihood of clicking on these links.     

\section{Conclusion}
Understanding the possible risks of embedded URLs within YouTube kids' videos is essential to provide a safe environment to children, without exposing them to inappropriate content. In this work, we studied the URLs embedded in comments on YouTube kids videos, focusing on their content topic, and the presence of malicious URLs. We collected a large-scale dataset of 3.7 million comments on YouTube kids' videos, and extracted 8,677 URLs embedded in the comments. We studied the users' interaction with such comments, and the potential risks of kids being exposed to inappropriate content or victims to some sort of malicious activities. Our findings highlight an increasing trend in inappropriate and malicious URLs within the comments, calling for increased awareness of such exposure and take measures to ensure children's safety from this exposure while on YouTube.

\BfPara{Acknowledgement} This work was supported by NRF grant 2016K1A1A2912757 (Global Research Lab) and a gift from NVIDIA.  S. Alshamrani was supported by a scholarship from the Saudi Arabian Cultural Mission.

\balance

\bibliographystyle{IEEEtran}
\bibliography{references}

\begin{thebibliography}{10}
\providecommand{\url}[1]{#1}
\csname url@samestyle\endcsname
\providecommand{\newblock}{\relax}
\providecommand{\bibinfo}[2]{#2}
\providecommand{\BIBentrySTDinterwordspacing}{\spaceskip=0pt\relax}
\providecommand{\BIBentryALTinterwordstretchfactor}{4}
\providecommand{\BIBentryALTinterwordspacing}{\spaceskip=\fontdimen2\font plus
\BIBentryALTinterwordstretchfactor\fontdimen3\font minus
  \fontdimen4\font\relax}
\providecommand{\BIBforeignlanguage}[2]{{%
\expandafter\ifx\csname l@#1\endcsname\relax
\typeout{** WARNING: IEEEtran.bst: No hyphenation pattern has been}%
\typeout{** loaded for the language `#1'. Using the pattern for}%
\typeout{** the default language instead.}%
\else
\language=\csname l@#1\endcsname
\fi
#2}}
\providecommand{\BIBdecl}{\relax}
\BIBdecl

\bibitem{GasserCML12}
U.~Gasser, S.~Cortesi, M.~M. Malik, and A.~Lee, ``Youth and digital media: From
  credibility to information quality,'' \emph{Berkman Center Research
  Publication}, no. 2012-1, 2012.

\bibitem{PewInternet2}
\BIBentryALTinterwordspacing
{Smith, Aaron and Toor, Skye and Kessel, Patric Van}. (2020) Many turn to
  youtube for children’s content, news, how-to lessons. [Online]. Available:
  \url{https://www.pewresearch.org/internet/2018/11/07/many-turn-to-youtube-for-childrens-content-news-how-to-lessons/}
\BIBentrySTDinterwordspacing

\bibitem{familyzone}
\BIBentryALTinterwordspacing
{FamilyZone}. (2020) Familyzone. [Online]. Available:
  \url{https://www.familyzone.com/au/families/blog/what-kids-did-online-2016}
\BIBentrySTDinterwordspacing

\bibitem{kaushalSBKP16}
R.~Kaushal, S.~Saha, P.~Bajaj, and P.~Kumaraguru, ``Kidstube: Detection,
  characterization and analysis of child unsafe content \& promoters on
  youtube,'' in \emph{the 14th Annual Conference on Privacy, Security and Trust
  (PST)}, 2016, pp. 157--164.

\bibitem{TahirASAZW19}
R.~Tahir, F.~Ahmed, H.~Saeed, S.~Ali, F.~Zaffar, and C.~Wilson, ``Bringing the
  kid back into youtube kids: Detecting inappropriate content on video
  streaming platforms,'' in \emph{IEEE/ACM International Conference on Advances
  in Social Networks Analysis and Mining}, 2019.

\bibitem{AlshamraniAAM20}
S.~Alshamrani, M.~Abuhamad, A.~Abusnaina, and D.~Mohaisen, ``Investigating
  online toxicity in users interactions with the mainstream media channels on
  youtube,'' in \emph{The 5th International Workshop on Mining Actionable
  Insights from Social Networks}, 2020, pp. 1--6.

\bibitem{MohaisenAM15}
A.~Mohaisen, O.~Alrawi, and M.~Mohaisen, ``Amal: High-fidelity, behavior-based
  automated malware analysis and classification,'' \emph{computers \&
  security}, vol.~52, pp. 251--266, 2015.

\bibitem{SpauldingPKM18}
J.~Spaulding, J.~Park, J.~Kim, and A.~Mohaisen, ``Proactive detection of
  algorithmically generated malicious domains,'' in \emph{2018 International
  Conference on Information Networking (ICOIN)}.\hskip 1em plus 0.5em minus
  0.4em\relax IEEE, 2018, pp. 21--24.

\bibitem{ThomasM14}
M.~Thomas and A.~Mohaisen, ``Kindred domains: detecting and clustering botnet
  domains using {DNS} traffic,'' in \emph{23rd International World Wide Web
  Conference, {WWW} '14, Seoul, Republic of Korea, April 7-11, 2014, Companion
  Volume}, C.~Chung, A.~Z. Broder, K.~Shim, and T.~Suel, Eds.\hskip 1em plus
  0.5em minus 0.4em\relax {ACM}, 2014, pp. 707--712.

\bibitem{ManadhataYRH14}
P.~K. Manadhata, S.~Yadav, P.~Rao, and W.~Horne, ``Detecting malicious domains
  via graph inference,'' in \emph{European Symposium on Research in Computer
  Security}, 2014, pp. 1--18.

\bibitem{Mohaisen15}
A.~Mohaisen, ``Towards automatic and lightweight detection and classification
  of malicious web contents,'' in \emph{Third {IEEE} Workshop on Hot Topics in
  Web Systems and Technologies, HotWeb 2015, Washington, DC, USA, November
  12-13, 2015}.\hskip 1em plus 0.5em minus 0.4em\relax {IEEE} Computer Society,
  2015, pp. 67--72.

\bibitem{KosbaMWTK14}
A.~E. Kosba, A.~Mohaisen, A.~G. West, T.~Tonn, and H.~K. Kim, ``{ADAM:}
  automated detection and attribution of malicious webpages,'' in
  \emph{Information Security Applications - 15th International Workshop, {WISA}
  2014, Jeju Island, Korea, August 25-27, 2014. Revised Selected Papers}, ser.
  Lecture Notes in Computer Science, K.~H. Rhee and J.~H. Yi, Eds., vol.
  8909.\hskip 1em plus 0.5em minus 0.4em\relax Springer, 2014, pp. 3--16.

\bibitem{WestM14}
A.~G. West and A.~Mohaisen, ``Metadata-driven threat classification of network
  endpoints appearing in malware,'' in \emph{Detection of Intrusions and
  Malware, and Vulnerability Assessment - 11th International Conference,
  {DIMVA} 2014, Egham, UK, July 10-11, 2014. Proceedings}, ser. Lecture Notes
  in Computer Science, S.~Dietrich, Ed., vol. 8550.\hskip 1em plus 0.5em minus
  0.4em\relax Springer, 2014, pp. 152--171.

\bibitem{ChoiAAWCNM19}
J.~Choi, A.~Abusnaina, A.~Anwar, A.~Wang, S.~Chen, D.~Nyang, and A.~Mohaisen,
  ``Honor among thieves: Towards understanding the dynamics and
  interdependencies in iot botnets,'' in \emph{2019 IEEE Conference on
  Dependable and Secure Computing (DSC)}.\hskip 1em plus 0.5em minus
  0.4em\relax IEEE, 2019, pp. 1--8.

\bibitem{SaadAM19}
M.~Saad, A.~Khormali, and A.~Mohaisen, ``Dine and dash: Static, dynamic, and
  economic analysis of in-browser cryptojacking,'' in \emph{2019 APWG Symposium
  on Electronic Crime Research (eCrime)}.\hskip 1em plus 0.5em minus
  0.4em\relax IEEE, 2019, pp. 1--12.

\bibitem{ZhangM0LX19}
M.~Zhang, W.~Meng, S.~Lee, B.~Lee, and X.~Xing, ``All your clicks belong to me:
  Investigating click interception on the web,'' in \emph{28th {USENIX}
  Security Symposium, {USENIX} Security 2019, Santa Clara, CA, USA, August
  14-16, 2019}, 2019, pp. 941--957.

\bibitem{EsheteVWZ13}
B.~Eshete, A.~Villafiorita, K.~Weldemariam, and M.~Zulkernine, ``{EINSPECT:}
  evolution-guided analysis and detection of malicious web pages,'' in
  \emph{37th Annual {IEEE} Computer Software and Applications Conference,
  {COMPSAC} 2013, Kyoto, Japan, July 22-26, 2013}, 2013, pp. 375--380.

\bibitem{Ranker}
\BIBentryALTinterwordspacing
{Ranker}. (2020) Ranker. [Online]. Available:
  \url{https://www.ranker.com/crowdranked-list/my-favorite-cartoons-of-all-time?ref=search}
\BIBentrySTDinterwordspacing

\bibitem{CommonSenseMedia}
\BIBentryALTinterwordspacing
{CommonSenseMedia}. (2020) Common sense media. [Online]. Available:
  \url{www.commonsensemedia.org}
\BIBentrySTDinterwordspacing

\bibitem{OGCK2011}
G.~S. O{\textquoteright}Keeffe, K.~Clarke-Pearson \emph{et~al.}, ``The impact
  of social media on children, adolescents, and families,'' \emph{Pediatrics},
  vol. 127, no.~4, pp. 800--804, 2011.

\bibitem{BerminghamCMOS09}
A.~Bermingham, M.~Conway, L.~McInerney, N.~O'Hare, and A.~F. Smeaton,
  ``Combining social network analysis and sentiment analysis to explore the
  potential for online radicalisation,'' in \emph{2009 International Conference
  on Advances in Social Network Analysis and Mining}, 2009, pp. 231--236.

\bibitem{EzpeletaIGMZ18}
E.~Ezpeleta, M.~Iturbe, I.~Garitano, I.~V. de~Mendizabal, and U.~Zurutuza, ``A
  mood analysis on youtube comments and a method for improved social spam
  detection,'' in \emph{Proceedings of 13th International Conference on Hybrid
  Artificial Intelligent Systems {HAIS} 2018, Spain,}, 2018, pp. 514--525.

\bibitem{CunhaCP19}
A.~A.~L. Cunha, M.~C. Costa, and M.~A.~C. Pacheco, ``Sentiment analysis of
  youtube video comments using deep neural networks,'' in \emph{Proceedings of
  the 18th International Conference in Artificial Intelligence and Soft
  Computing {ICAISC} 2019, Zakopane, Poland, Part {I}}, 2019, pp. 561--570.

\bibitem{PocheJWSVM17}
E.~Poch{\'{e}}, N.~Jha, G.~Williams, J.~Staten, M.~Vesper, and A.~Mahmoud,
  ``Analyzing user comments on youtube coding tutorial videos,'' in
  \emph{Proceedings of the 25th International Conference on Program
  Comprehension, {ICPC} 2017, Buenos Aires, Argentina}, 2017, pp. 196--206.

\bibitem{FigueiredoABG14}
F.~Figueiredo, J.~M. Almeida, F.~Benevenuto, and K.~P. Gummadi, ``Does content
  determine information popularity in social media?: a case study of youtube
  videos' content and their popularity,'' in \emph{{CHI} Conference on Human
  Factors in Computing Systems}, 2014, pp. 979--982.

\bibitem{imdb}
\BIBentryALTinterwordspacing
{IMDB}. (2020) Imdb. [Online]. Available: \url{www.imdb.com}
\BIBentrySTDinterwordspacing

\bibitem{FigueiredoBA11}
F.~Figueiredo, F.~Benevenuto, and J.~M. Almeida, ``The tube over time:
  characterizing popularity growth of youtube videos,'' in \emph{Proceedings of
  the Forth International Conference on Web Search and Web Data Mining, {WSDM}
  2011, Hong Kong, China}, 2011, pp. 745--754.

\bibitem{webshrinker}
Developers. (2020) Webshrinker. {https://www.webshrinker.com/}.

\bibitem{VirusTotal}
\BIBentryALTinterwordspacing
VirusTotal. (2020) {VirusTotal}. [Online]. Available:
  \url{https://www.virustotal.com/}
\BIBentrySTDinterwordspacing

\end{thebibliography}

\end{document}